\documentclass[nofootinbib,reprint,amsmath,amssymb,showkeys, showpacs,aps]{revtex4-1}
\usepackage{graphicx}
\usepackage[T1]{fontenc}
\usepackage{dcolumn}
\usepackage{xcolor,colortbl}
\usepackage{multirow}
\usepackage{bm}
\newcommand{\qm}[1]{``#1''}
\usepackage{hyperref}
\hypersetup{colorlinks, linkcolor={red},citecolor={blue},urlcolor={blue}}  
\newcommand\ChangeRT[1]{\noalign{\hrule height #1}}

\newcommand{\dd}[0]{{\rm d}}

\begin{document}

\title[Epicyclic frequencies in the equatorial plane around stationary and axially symmetric wormhole geometries]{Epicyclic frequencies in the equatorial plane around\\ stationary and axially symmetric wormhole geometries}

\author{Vittorio De Falco$^{1,2}$}\email{v.defalco@ssmeridionale.it}

\affiliation{$^1$ Scuola Superiore Meridionale,  Largo San Marcellino 10, 80138 Napoli, Italy,\\
$^2$ Istituto Nazionale di Fisica Nucleare, Sezione di Napoli, Complesso Universitario di Monte S. Angelo, Via Cintia Edificio 6, 80126 Napoli, Italy}

\date{\today}

\begin{abstract}
Epicyclic frequencies are usually observed in X-ray binaries and constitute a powerful astrophysical mean to probe the strong gravitational field around a compact object. We consider them in the equatorial plane around a general stationary and axially symmetric wormhole. We first search for the wormholes' existence, distinguishing them from a Kerr black hole. Once there will be available observational data on wormholes, we present a strategy to reconstruct the related metrics. Finally, we discuss the implications of our approach and outline possible future perspectives.
\end{abstract}

\maketitle
\section{Introduction}
\label{sec:intro}
A wormhole (WH) is an exotic compact object, characterized by a non-trivial topology featuring no horizons and physical singularities. Furthermore, it presents a traversable bridge, dubbed WH neck, connecting two distinct universes or two different regions of the same spacetime \cite{Visser1995}. This topic is frequently studied both in General Relativity (GR) and in Alternative/Extended Theories of gravity, where the related works can be classified in two macro-research areas: (1) proposing new WH solutions in different gravity frameworks by employing disparate mathematical methods (see e.g., Refs. \cite{Visser1989,Anchordoqui:2000ut,Bahamonde:2016jqq,Capozziello2020}); (2) providing original astrophysical strategies based on the current or near-future observational data to look for the detection of WH existence (see e.g., Refs. \cite{Cardoso2016,Dai2019,Dalui2019,Defalco2020WH,DeFalco2021,DeFalco2021WF}). 

Since these exotic objects have never been observed so far, it could be related to the fact that probably there exist particular WHs, which perfectly reproduce all observational properties of a BH with arbitrary high-accuracy, known also in the literature as \emph{black hole (BH) mimickers} \cite{Cardoso2019}. In order to reveal their existence, it would be very useful to provide tests of gravity in strong field regime.

To this purpose, a helpful astrophysical tool of investigation is represented by the \emph{epicyclic frequencies}. The term \qm{epicyclic} derives from the Greek and it means \emph{beyond the circle}. Indeed, such frequencies $\left\{\nu_r,\nu_\varphi,\nu_\theta\right\}$ are physically obtained by linearly perturbing the motion of a test particle in a circular orbit along the radial, azimuthal, and polar directions, respectively. The epicyclic frequencies entail several advantages, because they closely depend on the underlying geometrical background, are produced in strong field regime, and are frequently found in BH systems \cite{Motta2016,Ingram2020}.

In the literature, it is possible to find already some works on epicyclic frequencies applied to WHs, whose objectives are: (1) understanding the behaviour of a test gyroscope moving towards a Teo rotating traversable WH \cite{Chakraborty2017}; (2) analysis of quasi-periodic oscillations (QPOs) from an accretion disk around Teo rotating traversable WHs  \cite{Deligianni2021}; (3) testing observationally the presence of BH mimicker solutions via QPOs \cite{Jiang2021}; (4) investigations of the epicyclic frequencies around Simpson-Visser regular BHs and WHs \cite{Stuchlik2021a}; (5) application of epicyclic orbits in the field of Einstein-Dirac-Maxwell traversable WHs to the QPOs observed in microquasars and active galactic nuclei \cite{Stuchlik2021b}; (6) studies on the epicyclic frequencies around traversable phantom WHs in Rastall gravity \cite{Javed2022}.

In a previous work, we have studied the epicyclic frequencies in general static and spherically symmetric WH spacetimes, where we have showed the strategy to disentangle between a BH and a WH, and how to reconstruct a WH solution once they will be detected \cite{DeFalco2021EF}. In this work, we aim at extending the aforementioned approach to general stationary and axially symmetric WH geometries. Therefore, the paper is structured as follows: in Sec. \ref{sec:epicyclic-frequencies-Teo-WHs} we describe the epicyclic frequencies around stationary and axially symmetric WHs; in Sec. \ref{sec:applications} we first describe the procedure to detect possible metric deviations from a Kerr BH and then explain how it is possible to reconstruct the WH solution from the observational data; finally in Sec. \ref{sec:end} we draw the conclusions. 

\section{Epicyclic frequencies in stationary and axially symmetric wormholes}
\label{sec:epicyclic-frequencies-Teo-WHs}
In this section, we first introduce general stationary, axially symmetric, and traversable WH geometries described by the Teo-like metric (see Sec. \ref{sec:Teo-WHs}) and then we present the formulas of the epicyclic frequencies in the equatorial plane of such spacetimes (see Sec. \ref{sec:epicyclic-frequencies}). 

From this section onward, we use geometrical units $G=c=1$ and the distances will be measured in units of $M$, being the total mass-energy of the considered compact object generating the underlying gravitational field.

\subsection{Teo-like wormholes}
\label{sec:Teo-WHs}
General stationary, axially symmetric, and traversable WHs can be described in spherical-like coordinates $(t,r,\theta,\varphi)$ employing the following Teo-like metric \cite{Teo1998} 
\begin{align} \label{eq:Teo_WH}	
\dd s^2&=-N^2(r,\theta) \dd t^2 + \frac{\dd r^2}{1-\frac{b(r,\theta)}{r}}\notag\\
&+r^2K^2(r,\theta)\biggr{[}\dd \theta^2+\sin^2\theta\biggr{(}\dd \varphi-\omega(r,\theta)\dd t\biggr{)}^2\biggr{]},
\end{align}
where $N(r,\theta),b(r,\theta),K(r,\theta),\omega(r,\theta)$ are four unknown functions, which determine the WH spacetime. In particular, we have: $N(r,\theta)$ is the redshift function and describes the time properties of the WH; $b(r,\theta)$ is the shape function, delineating the WH form when it is embedded in an Euclidean space\footnote{In the case of Morris-Thorne (static and spherically symmetric) WHs \cite{Morris1988}, these solutions are embedded in a three-dimensional Euclidean space, since the metric is invariant with respect to the $\theta$ coordinate. Instead, Teo-like WHs should be embedded in a four-dimensional Euclidean space, after having fixed a time instant, which does not spoil the final WH shapes due to their stationary and axial symmetry properties.};
$K(r,\theta)$ is the proper radial distance factor, which permits to define the proper radial distance $R=rK(r,\theta)$ from the origin of the coordinate system, endowed with the property to have $\partial R/\partial r>0$; $\omega(r,\theta)$ is the rotational function devoted to characterize the frame-dragging effect around the WH. Equation \eqref{eq:Teo_WH} reduces to the Morris-Thorne metric \cite{Morris1988} in the limit of zero rotation (i.e., $\omega(r,\theta)\to0$) and spherical symmetry, which in formulas translates in requiring
\begin{equation} 
N(r,\theta)\to e^{\Phi(r)},\quad b(r,\theta)\to b(r),\quad K(r,\theta)\to1.
\end{equation} 

Such WHs must fulfill the following properties \cite{Teo1998}:
\begin{enumerate}
\item for having no horizons the $\theta$-derivatives of $N(r,\theta),\ b(r,\theta),\ K(r,\theta)$ evaluated in $\theta=0,\ \pi$ have to vanish on the rotation axis;
\item defined $r_0>0$ as the WH throat, no essential singularities occur if $N(r,\theta),\ b(r,\theta),\ K(r,\theta),\ \omega(r,\theta)$ are smooth functions everywhere finite for $r\ge r_0$;
\item the shape function fulfills: $b\le r$, $\partial_\theta b(r_0,\theta)=0$ for all $\theta\in[0,\pi]$, $b>r\partial_r b $ (flaring-out condition);
\item asymptotically flatness, i.e., for $r\to\infty$, we have $N\to1,\frac{b}{r}\to0,\ K\to1,\ \omega\to0$;
\item the metric \eqref{eq:Teo_WH} is valid both in GR and Extended/Alternative theories of gravity. It generally depends on the Arnowitt-Deser-Misner (ADM) mass (or total mass-energy of the system contained in the whole spacetime \cite{Visser1995}) $M$, the (dimensionless Kerr spin-like) total angular momentum $a$, and sometimes from other parameters, which come from the gravity theory to which it belongs and also the employed stress-energy tensor to construct it; 
\item the traversability is achieved by either resorting to quantum mechanical effects, produced by \emph{ad hoc} exotic stress-energy tensors (see e.g., Refs. \cite{Hochberg1997,Bronnikov2013,Garattini2019}), or topological arguments, based on standard and (gravitational) curvature fluid stress-energy tensors (see e.g., Refs. \cite{Lobo2009,Harko2013,Capozziello2012}). The former approach is generally employed in GR, whereas the latter in modified gravity frameworks, presenting more degrees of freedom with respect to GR.
\end{enumerate}

\subsection{Epicyclic frequencies in the equatorial plane}
\label{sec:epicyclic-frequencies}
The epicyclic frequencies $\{\nu_r,\nu_\varphi,\nu_\theta\}$ are normally calculated in terms of the epicyclic angular velocities $\{\Omega_r=2\pi \nu_r,\Omega_\varphi=2\pi \nu_\varphi,\Omega_\theta=2\pi \nu_\theta\}$, whose explicit formulas can be obtained in the equatorial plane $\theta=\pi/2$ by exploiting one of the following equivalent strategies:
\begin{itemize}
\item employing the \emph{conserved specific energy $\mathcal{E}$ and angular momentum $\ell$ along the test particle trajectory}, it is possible to write the following expressions 
\begin{equation}
\dot{t}=\dot{t}(\mathcal{E},\ell),\qquad \dot{\varphi}=\dot{\varphi}(\mathcal{E},\ell),
\end{equation}
where dot stays for the derivative with respect to an affine parameter along the test particle trajectory. Using the normalization condition for timelike four-velocities $g_{\mu\nu}\dot{x}^\mu\dot{x}^\nu=-1$, we have
\begin{equation}
g_{rr}\dot{r}^2+g_{\theta\theta}\dot{\theta}^2=\mathcal{V}_{\rm eff}(r,\theta,\mathcal{E},\ell).
\end{equation}
For stable circular orbits in the equatorial plane we have $\dot{r}=\dot{\theta}=0$, which implies $\mathcal{V}_{\rm eff}=0$, whereas $\ddot{r}=0$ and $\ddot{\theta}=0$ entails
\begin{equation}
\partial_r\mathcal{V}_{\rm eff}=0,\qquad \partial_\theta\mathcal{V}_{\rm eff}=0.
\end{equation}
We then have (see Sec. 10.3.2 in Ref. \cite{Bambi2018})
\begin{subequations} \label{eq:eqEFs}
\begin{align}
\left(\frac{\dd r}{\dd t}\right)^2&=\frac{1}{g_{rr}\dot{t}^2}\mathcal{V}_{\rm eff},\\
\left(\frac{\dd \theta}{\dd t}\right)^2&=\frac{1}{g_{\theta\theta}\dot{t}^2}\mathcal{V}_{\rm eff}.
\end{align}
\end{subequations}
We derive Eqs. \eqref{eq:eqEFs} with respect to the coordinate time $t$ and then consider $r=r_0+\delta r$ and $\theta=\pi/2+\delta \theta$, where $\delta r$ and $\delta\theta$ are small perturbations. We linearize the system and obtain harmonic oscillator equations, which provides the expressions of $\Omega_r,\Omega_\theta,\Omega_\varphi$ in terms of the metric \cite{Bambi2018,Turimov2022};
\item starting from the timelike geodesic equations, we can employ the \emph{relativity of observer splitting formalism} \cite{Jantzen1992,Bini1997a,Bini1997b,Defalco2018}\footnote{This technique permits to clearly distinguish between gravitational and inertial contributions. It encompasses a direct connection with the classical description and allows us to reveal the physics behind the symbols we algebraically manipulate.} and the zero angular momentum observers (ZAMOs). Therefore, the test particle's position $(r,\varphi)$ is expressed in spherical-like coordinates, whereas its spatial velocity vector $\boldsymbol{\nu}$ is split in the ZAMO frame $\{\boldsymbol{e}_{\hat t},\boldsymbol{e}_{\hat r},\boldsymbol{e}_{\hat \theta},\boldsymbol{e}_{\hat \varphi}\}$ through $(\nu,\alpha)$, where $\nu=||\boldsymbol{\nu}||$ is the magnitude of the spatial velocity and $\alpha$ is the azimuthal angle of the vector $\boldsymbol{\nu}$ in the $\boldsymbol{e}_{\hat r}-\boldsymbol{e}_{\hat \varphi}$ plane measured clockwise from the positive $\boldsymbol{e}_{\hat \varphi}$ direction. We finally obtain
\begin{subequations} \label{eq:dynamical-system}
\begin{align}
\frac{\dd \nu}{\dd t}&=f_1(\nu,\alpha,r),\\
\frac{\dd \alpha}{\dd t}&=f_2(\nu,\alpha,r),\\
\frac{\dd r}{\dd t}&=f_3(\nu,\alpha,r).
\end{align}
\end{subequations}
We perturb the above dynamical system around a stable circular orbit of radius $r_0$ endowed with Keplerian velocity (i.e., $\alpha_0=0$ and $\nu_0=\nu_K(r)$\footnote{To calculate $\nu_K$, we can employ $\nu_K=r\Omega_K(r,\theta)$, with $\Omega_k(r,\theta)$ being the Keplerian angular velocity (cf. Eq. \eqref{eq:Omega_f}.}) via a small parameter $\varepsilon\ll1$, namely
\begin{equation}
\nu=\nu_K+\varepsilon \nu_1, \qquad \alpha=\varepsilon \alpha_1,\qquad r=r_0+\varepsilon r_1,
\end{equation}
Linearizing the dynamical system \eqref{eq:dynamical-system}, we have 
\begin{subequations} \label{eq:dynamical-system2}
\begin{align}
\frac{\dd \nu_1}{\dd t}&=\tilde{f}_1(\alpha_1,r_0), \label{eq:nu_1}\\
\frac{\dd \alpha_1}{\dd t}&=\tilde{f}_2(\nu_1,r_1,r_0),\\
\frac{\dd r_1}{\dd t}&=\tilde{f}_3(\alpha_1,r_0).\label{eq:r_1}
\end{align}
\end{subequations}
Now, we consider
\begin{align}
\frac{\dd^2 \alpha_1}{\dd t^2}&=\frac{\partial \tilde{f}_2}{\partial \nu_1}\frac{\dd \nu_1}{\dd t}+\frac{\partial \tilde{f}_2}{\partial r_1}\frac{\dd r_1}{\dd t},
\end{align}
and substituting Eqs. \eqref{eq:nu_1} and \eqref{eq:r_1}, it leads to the harmonic oscillator equation
\begin{align}
\frac{\dd^2 \alpha_1}{\dd t^2}+\Omega_r^2\alpha_1=0,
\end{align}
where we obtain the explicit expression of $\Omega_r$. 

The azimuthal epicyclic frequency is calculated through the Keplerian angular velocity $\Omega_K$, i.e.,
\begin{equation}
\Omega_\varphi\equiv \Omega_K=\frac{\dd \varphi}{\dd t}.
\end{equation}
For determining $\Omega_\theta$, we should first introduce the polar angle $\psi$ in the ZAMO frame, measured from the $\boldsymbol{e}_{\hat \theta}$ direction, and then following a similar procedure outlined above for determining $\Omega_r$ \cite{DeFalco20183D,Bakala2019}. 
\end{itemize}
The epicyclic angular velocities' formulas, evaluated at the angle $\theta=\pi/2$ and radius $r=r_0$, are \cite{Abramowicz2005}
\begin{subequations} 
\begin{align}
\Omega_\varphi&=\frac{-\partial_r g_{t\varphi}\pm\sqrt{(\partial_r g_{t\varphi})^2-(\partial_r g_{tt})(\partial_r g_{\varphi\varphi})}}{\partial_r g_{\varphi\varphi}}, \label{eq:Omega_f}\\
\Omega_r^2&=\frac{(g_{tt}+\Omega_\varphi g_{t\varphi})^2}{2g_{rr}}\biggr{[}\partial_{rr}^2\biggr{(}\frac{g_{\varphi\varphi}}{Y}\biggr{)}+2\ell \partial_{rr}^2\biggr{(}\frac{g_{t\varphi}}{Y}\biggr{)}\notag\\
& \hspace{3cm}+\ell^2\partial_{rr}^2\biggr{(}\frac{g_{tt}}{Y}\biggr{)}\biggr{]},\label{eq:Omega_r}\\
\Omega_\theta^2&=\frac{(g_{tt}+\Omega_\varphi g_{t\varphi})^2}{2g_{\theta\theta}}\biggr{[}\partial_{\theta\theta}^2\biggr{(}\frac{g_{\varphi\varphi}}{Y}\biggr{)}+2\ell \partial_{\theta\theta}^2\biggr{(}\frac{g_{t\varphi}}{Y}\biggr{)}\notag\\
& \hspace{3cm}+\ell^2\partial_{\theta\theta}^2\biggr{(}\frac{g_{tt}}{Y}\biggr{)}\biggr{]}, \label{eq:Omega_th}
\end{align}
\end{subequations}
where
\begin{subequations} 
\begin{align}
Y&=g_{tt}g_{\varphi\varphi}-g_{t\varphi}^2,\\
\ell&=-\frac{g_{t\varphi}+\Omega_\varphi g_{\varphi\varphi}}{g_{tt}+\Omega_\varphi g_{t\varphi}}.
\end{align}
\end{subequations}

\section{Searching for wormhole's existence and metric reconstruction}
\label{sec:applications}
The epicyclic frequencies can be normally found in several \emph{X-ray binaries}, composed by a BH (or a neutron star) and a companion donor star. These systems are characterized by the presence of an accretion disk, strongly emitting in the X-ray energy band, and by frequently flux variabilities on short timescales \cite{Lewin1997}. The latter effects are studied within the Fourier analysis via \emph{power-density spectra}, which features very fast aperiodic and quasi-periodic variabilities showing (generally) the existence of narrow peaks with a distinct centroid frequencies, also known as QPOs (see Refs. \cite{Motta2016,Ingram2020}, for reviews). Although their origin is still not clear, the cause of their production is associated with the strong gravity's interaction with the motion of the matter around massive compact objects. \emph{QPO models share an extensive use of the epicyclic frequencies framed within different theoretical patterns} \cite{Ingram2020}. Therefore, once we detect them, we should choose the appropriate theoretical model in order to infer the right values of the epicyclic frequencies. This represents the main criticality of this procedure, because sometimes it could be difficult to pinpoint the right QPO model, or more than one model could be employed (producing a model-degeneracy), or in the worst case no model could be exploited for the available data \cite{Motta2016,Ingram2020}.

We stress that this is a promising approach for the availability of actual and also near-future more accurate observational data (see e.g., Refs. \cite{Remillard2006,Feroci2016,Zhang2016,Soffitta2013}). In this section, we first describe how to distinguish between a BH and the presence of a WH (see Sec. \ref{sec:WHdet}). If a WH is detected, we propose a methodology to reconstruct the related solution from the observations (see Sec. \ref{sec:WHrec}).

\subsection{Method to distinguish between a Teo-like wormhole and a Kerr black hole}
\label{sec:WHdet}
The technique to distinguish between a Kerr BH and a Teo-like WH consists in detecting metric-departures from the BH geometries in GR. Therefore, if we are able to fit the data on epicyclic frequencies via the Kerr model, then no WH is present; otherwise, a WH may exist. 

Since there are no epicyclic frequencies' data associated to WHs, we select some WH solutions from the literature. We would like to clarify that differently from the static and spherically symmetric case, where a WH solution can be found relatively easy, since only two unknown functions (i.e., $g_{tt}(r)$ and $g_{rr}(r)$) must be determined, in the stationary and axially symmetric situation more functions and a dependence also from the polar angle $\theta$ are involved (i.e., $g_{tt}(r,\theta)$, $g_{rr}(r,\theta)$, $g_{\varphi\varphi}(r,\theta)$, $g_{\theta\theta}(r,\theta)$, and $g_{t\varphi}(t,\theta)$). Therefore, some ansätze are generally invoked in order to restrict the functional space of the solutions. The proposed WH geometries, reported in Table \ref{tab:Table1}, are all exact solutions of the field equations in GR, obtained by resorting to different models of exotic fluids.
\renewcommand{\arraystretch}{2.2}
\begin{table*}[th!]
\begin{center}
\caption{\label{tab:Table1} Some examples of WH solutions in GR (obtained by resorting to different exotic stress-energy tensors) are displayed. We show the general expression of each metric component in the equatorial plane $\theta=\pi/2$. For all WH solutions, we set $a=0.3$, $r_0=M$, and in the column \qm{\emph{PARAM.}}, we assign numerical values to the free parameters. In the first row ($\#0$), we report the Kerr BH solution for comparing it with the other WH solutions (from $\#1$ to $\#10$).\\}
\scalebox{1.14}{
\begin{tabular}{|@{} c@{} |@{} c @{}| @{}c @{}|@{} c @{}|@{} c @{}|@{} c @{}|@{} c @{}|} 
\ChangeRT{1pt}
$\quad\boldsymbol{\#}\quad$ &$\quad\boldsymbol{g_{tt}}\quad$ & $\quad\boldsymbol{g_{rr}}\quad$  &  $\quad\boldsymbol{g_{\varphi\varphi}}\quad$  &  $\quad\boldsymbol{g_{t\varphi}}\quad$ &\quad $ \boldsymbol{{\rm PARAM.}}$ \quad & \quad {\bf Ref.} \quad  \\
\hline
\hline
\rowcolor{lightgray} 0 & $-\left(1-\dfrac{2M}{r}\right)$ & $\dfrac{r^2}{r^2-2Mr+a^2}$ & $r^2+a^2+\dfrac{2Ma^2}{r}$ & $-\dfrac{2Ma}{r}$ & --&\cite{Wiltshire2009}\\
\hline
1\footnote{$\Delta_1=(r-l_1)^2+(l_0^2-l_1^2)$, $\Delta_2=\frac{a}{2(r-l_1)}$ with $l_0^2>l_1^2> 0$.} & $-1$ & $\dfrac{(r-l_1)^2}{\Delta_1}$ & $\Delta_1-\Delta_2^2$ & $-\Delta_2$ & $l_0=1.1M,l_1=M$ & \cite{Miranda2014}\\
\hline
2\footnote{$\delta=\frac{\lg\beta}{\lg(1-\beta)}$ with $\beta=r_h/r_0>1$.} & $-\left(1-\dfrac{r_h}{r}\right)^{1+\delta}$ & $\left\{1-\dfrac{r_h}{r}\left[1+\left(1-\dfrac{r_h}{r}\right)^{1-\delta}\right]\right\}^{-1}$ & $r^2$ & $-\dfrac{2Ma}{r}$ & $r_h=0.4r_0$ & \cite{Abdujabbarov2009}\\
\hline
3 & $-1+\dfrac{4M^2a^2}{r^4}$ & $\left(1-\dfrac{r_0^2}{r^2}\right)^{-1}$ & $r^2$ & $-\dfrac{2Ma}{r}$ & -- &\cite{Kim2005}\\
\hline
4 & $-e^{-\frac{r_0}{r}}+\dfrac{4M^2a^2}{r^4}$ & $\left(1-\dfrac{r_0}{r}\right)^{-1}$ & $r^2$ & $-\dfrac{2Ma}{r}$ & --& \cite{Harko2009}\\
\hline
5 & $-e^{-\frac{r_0}{r}}+\dfrac{4M^2a^2}{r^4}$ & $\left(1-\dfrac{r_0^2}{r^2}\right)^{-1}$ & $r^2$ & $-\dfrac{2Ma}{r}$ & --& \cite{Harko2009}\\
\hline
6 & $-e^{-\frac{r_0}{r}}+\dfrac{4M^2a^2}{r^4}$ & $\left(1-\dfrac{\sqrt{r_0r}}{r}\right)^{-1}$ & $r^2$ & $-\dfrac{2Ma}{r}$ & --& \cite{Harko2009}\\
\hline
7\footnote{$0<\gamma<1$.} & $-e^{-\frac{r_0}{r}}+\dfrac{4M^2a^2}{r^4}$ & $\left[1-\dfrac{r_0+\gamma r_0\left(1-\dfrac{r_0}{r}\right)}{r}\right]^{-1}$ & $r^2$ & $-\dfrac{2Ma}{r}$ & $\gamma=0.5$& \cite{Harko2009}\\
\hline
8\footnote{$\Sigma=r^2+h^2$, $\Delta=\Sigma+a^2-2M\sqrt{\Sigma}$, $C=(\Sigma+a^2)^2-a^2\Delta $.} & $-\left(1-\dfrac{2M}{\sqrt{\Sigma}}\right)$ & $\dfrac{\Sigma}{\Delta}$ & $\dfrac{C}{\Sigma}$ & $-\dfrac{2Ma}{\sqrt{\Sigma}}$ & $h=2$& \cite{Mazza2021}\\
\hline
9\footnote{$A=1-\frac{2M}{r+1.5r_0}$, $B=2M-\frac{1.5r_0(r-1.5r_0)}{r}$, $f=r^2(1-A)$, $\Phi=r^2A+a^2$, $\Psi=r^2A+a^2$, $\rho=(r^2+a^2)^2-a^2\Phi$ with $r_0>M$.} & $-\left(1-\dfrac{f}{r^2}\right)$ & $\dfrac{r^2A}{\Psi}\left(1-\dfrac{B}{r}\right)^{-1}$ & $\dfrac{\rho}{r^2}$ & $-a\dfrac{f}{r^2}$ &--& \cite{Azreg2016}\\
\hline
10 & $-1+\dfrac{4M^2a^2}{r^4}$ & $\left(1-\dfrac{r_0}{r}\right)^{-1}$ & $r^2$ & $-\dfrac{2Ma}{r}$ &--& \cite{Teo1998}\\
\ChangeRT{1pt}
\end{tabular}}
\end{center}
\end{table*}

Once we have fixed all free parameters of each WH solution (see Sec. \ref{sec:remark_sim}, where we provide more details on the employed methodology and the displayed simulations), in Table \ref{tab:Table2} we calculate the related WH throat $r_0$, innermost stable circular orbit (ISCO) radius $r_{\rm ISCO}$, and epicyclic frequencies (in the equatorial plane) $\Omega_\varphi$ and $\Omega_r$. Regarding the ISCO radius, it can be computed via the radial geodesic equation \cite{Chandrasekhar1983}, which could be very demanding for our WH solutions (similarly as it is done in the Kerr metric). An alternative and simpler manner to calculate $r_{\rm ISCO}$ can be achieved by determining the minimum value for which the radial epicyclic angular velocity $\Omega_r$ is defined. The knowledge of the ISCO radius is very important, because it permits to preliminarily understand the geometrical properties of a WH spacetime. 
\renewcommand{\arraystretch}{2.2}
\begin{table*}[th!]
\begin{center}
\caption{\label{tab:Table2} Each row corresponds to the BH/WH solution of Table \ref{tab:Table1}. In the first row, $r_0$ is the Kerr event horizon radius.\\}
\scalebox{0.63}{
\begin{tabular}{|@{} c@{} |@{} c @{}| @{}c @{}|@{} c @{}|@{} c @{}|} 
\ChangeRT{1pt}
$\quad\boldsymbol{\#}\quad$ &$\quad\boldsymbol{r_0}\quad$ & $\quad\boldsymbol{r_{\rm ISCO}}\quad$  &  $\quad\boldsymbol{\Omega_\varphi}\quad$  &  $\quad\boldsymbol{\Omega_r}\quad$ \\
\hline
\hline
\rowcolor{lightgray} 0 & $1.95M$  & $4.98M$ & $\dfrac{r^{3/2}-0.3}{r^3-0.1}$ & $\dfrac{-0.08 r^{3/2}+0.87 r^{5/2}-1.5 r^{7/2}+14. r^{9/2}-6.7 r^{11/2}-3.6 r^{13/2}+1.8 r^{15/2}}{r^2 ((r-2) r+0.1)^2 \left(r^3-0.1\right)^2}$\\
\rowcolor{lightgray}& & & &$\dfrac{+r^9-10 r^8+27.9 r^7-25.7 r^6+5.9 r^5-3.5 r^4-1.68 r^3+0.08 r^2}{r^2 ((r-2) r+0.1)^2 \left(r^3-0.1\right)^2}$ \\
\hline
1  &  $M$  &  $M$ & $\dfrac{0.5 \sqrt{2.24 r^{1.21} (r-0.4)^{1.79}+0.36}-0.3}{r^3}$ & $\dfrac{r (r (r (r (r (r (r (r (r (r (r (r (r (r (r (r ((0.02 r-0.4) r+3.5)-18.5)+70)-198)+435)}{(r-1)^{20} ((r-2) r+1.2)^2}$\\
& & & & $\dfrac{-758)+1062)-1206)+1112)-831)+500)-238)+88.3)-24.5)+4.8)-0.6)}{(r-1)^{20} ((r-2) r+1.2)^2}$\\
\hline
2\footnote{$k_2=r-0.4$, $k_3=0.4+ 2.2k_2^1.8r^{1.2}$, $k_4=2.2 r^{1.2}k_2^{1.8}+0.4$.} & $M$ & NO & $0$ & $\dfrac{0.3 \left(k_2^{1.8}-0.4 r^{0.8}\right) \big{[}r^{1.2} \left(2.2 \sqrt{k_2} k_3^{2.8}+\sqrt{k_4} k_2^{1.8}- k_2^{2.8}-k_2^{1.8}+0.1 k_2^{0.8}\right)}{k_2^{0.8} r^{7.} \left(r^{1.2} k_2^{2.8}+0.4\right)^3}$ \\
& & & & $\dfrac{+r^{3.6} \left(7.3 \sqrt{k_3} k_2^{7.4}+\left(3.3\sqrt{k_4}-0.7\right) k_2^{8.4}-6 k_2^{7.4}-1.7 k_2^{6.4}\right)}{k_2^{0.8} r^{7} \left(r^{1.2} k_2^{2.8}+0.4\right)^3}$\\
& & & & $\dfrac{+r^{2.4} \left(4.7 \sqrt{k_3} k_2^{5.6}+5.3 \sqrt{k_4} k_2^{4.6}-2 k_2^{5.6}-5. k_2^{4.6}\right)}{k_2^{0.8} r^{7} \left(r^{1.2} k_2^{2.8}+0.4\right)^3}$\\
& & & & $\dfrac{+\left(2. k_2^{10.2}-3 k_2^{9.2}\right) r^{4.8}+0.3 \sqrt{k_4}-0.2\big{]}}{k_2^{0.8} r^{7} \left(r^{1.2} k_2^{2.8}+0.4\right)^3}$\\
\hline
3  & $M$ & NO & $\dfrac{0.6}{r^3}$  & $0$\\
\hline
4\footnote{$k_5=4 e^{-2/r} r^3+3$.}  & $M$ & NO & $\dfrac{0.5 \sqrt{k_5}-0.3}{r^3}$  & $\dfrac{0.05 e^{-2/r} (r-1) \left(22(r-2) r^{10}+e^{2/r} \left(80 \sqrt{k_5}+r \left(20\sqrt{k_5}-36\right)-216\right) r^7+e^{4/r} \left(65 \sqrt{k_5}-117\right) r^4\right)}{r^{15}}$\\
\hline
5  & $M$ & NO & $\dfrac{0.5 \sqrt{4 e^{-2/r} r^3+3}-0.3}{r^3}$ & $\dfrac{e^{-2/r} \left(r^2-1\right) \left((r-2) r^{10}+e^{2/r} \left(r \left(0.9 \sqrt{4 e^{-2/r} r^3+3}-1.6\right)+3.6 \sqrt{k_5}-9.7\right) r^7+e^{4/r} \left(2.916 \sqrt{k_5}-5.2\right) r^4\right)}{r^{16}}$\\
\hline
6  & $M$ & NO & $\dfrac{0.5 \sqrt{k_5}-0.3}{r^3}$& $-\dfrac{e^{-2/r} \left(\frac{1}{\sqrt{r}}-1.\right) \left((r-2) r^{10}+e^{2/r} \left(r \left(0.9 \sqrt{k_5}-1.62\right)+3.6 \sqrt{k_5}-9.72\right) r^7+e^{4/r} \left(2.916 \sqrt{k_5}-5.2\right) r^4\right)}{r^{14}}$\\
\hline
7 & $M$ & NO &  $\dfrac{0.5 \sqrt{4 e^{-2/r} r^3+3.24}-0.3}{r^3}$ & $\dfrac{0.05 e^{-2/r} ((r-1.5) r+0.5) \left(22(r-2) r^{10}+e^{2/r} \left(80 \sqrt{k_5}+r \left(20 \sqrt{k_5}-36\right)-216\right) r^7+e^{4/r} \left(65 \sqrt{k_5}-117\right) r^4\right)}{r^{16}}$\\
\hline
8\footnote{$k_1=r^2+4$, $k_6=k_1^{49/4} \left(\sqrt{k} k_1^2-0.1 r^2\right)^2 \left(\sqrt{k_1} k_1^2-2 k_1^2\right)^3$, $k_7=-12r^{42}-1090 r^{40}-46932 r^{38}-10^6 r^{36}-2\times 10^7 r^{34}-4\times 10^8 r^{32}-4\times 10^9 r^{30}-4\times 10^{10} r^{28}-3\times 10^{11} r^{26}-2\times 10^{12} r^{24}-8\times 10^{12} r^{22}-3\times 10^{13} r^{20}-10^{14} r^{18}-3\times 10^{14} r^{16}-8\times 10^{14} r^{14}-2\times 10^{15} r^{12}-3\times 10^{15} r^{10}-3\times 10^{15} r^8-3\times 10^{15} r^6-2\times 10^{15} r^4-8\times 10^{14} r^2-2\times 10^{14}$, $k_8=r^{42}+132 r^{40}+7260 r^{38}+235593 r^{36}+5\times 10^6 r^{34}+8\times 10^7 r^{32}+10^9 r^{30}+10^{10} r^{28}+8\times 10^{10} r^{26}+5\times 10^{11} r^{24}+3\times 10^{12} r^{22}+10^{13} r^{20}+5\times 10^{13} r^{18}+10^{14} r^{16}+3\times 10^{14} r^{14}+7\times 10^{14} r^{12}+10^{15} r^{10}+10^{15} r^8+10^{15} r^6+9\times 10^{14} r^4+4\times 10^{14} r^2+7\times 10^{13}$, $k_9=2r^{42}+152 r^{40}+6070 r^{38}+152903 r^{36}+3\times 10^6 r^{34}+4\times 10^7 r^{32}+4\times 10^8 r^{30}+3\times 10^9 r^{28}+2\times 10^{10} r^{26}+10^{11} r^{24}+6\times 10^{11} r^{22}+2\times 10^{12} r^{20}+6\times 10^{12} r^{18}+2\times 10^{13} r^{16}+3\times 10^{13} r^{14}+6\times 10^{13} r^{12}+7\times 10^{13} r^{10}+7\times 10^{13} r^8+5\times 10^{13} r^6+2\times 10^{13} r^4+5\times 10^{12} r^2+3\times 10^{11}$, $k_{10}=-7r^{38}-578 r^{36}-22021 r^{34}-530180 r^{32}-9\times 10^6 r^{30}-10^8 r^{28}-10^9 r^{26}-9\times 10^9 r^{24}-6\times 10^{10} r^{22}-3\times 10^{11} r^{20}-2\times 10^{12} r^{18}-5\times 10^{12} r^{16}-2\times 10^{13} r^{14}-4\times 10^{13} r^{12}-8\times 10^{13} r^{10}-10^{14} r^8-2\times 10^{14} r^6-2\times 10^{14} r^4-10^{14} r^2-5\times 10^{13}$.} & $0$ & $4.56M$ & $\dfrac{k_1^2 \left(\sqrt[4]{k_1} \sqrt{k_1}-0.3\right)}{\left(48\sqrt{k_1}-0.72\right) r^2+\sqrt{k_1}(64 +r^6)+\left(12\sqrt{k_1}-0.09\right) r^4-1.44}$ & $\dfrac{r^2 [k_1^{3/4}k_7 +k_1^{5/4} k_8+\sqrt{k_1} k_9- 10^{13} r^2+r^4 k_{10}-5\times 10^{11}]}{k_6}$ \\
\hline
9\footnote{$k_{11}=r^{27/2} (r (r (r (r (r (r (r (r (r (r (r (r (r (r (r (r (r (r (r (r (r (r (r (r (r (r (r+21)+200)+1100)+4400)+12000)+24000)+33000)+33000)+23000)+14000)+11000)+12000)+7700)+1900)+130)+1100)+920)-45)-260)-18)+34)-16)-19)-5)-0.2)+0.04)-0.001)$, $k_{12}=-2\times10^{-53} r^{3/2}-6\times10^{-37} r^{5/2}-5\times10^{-36} r^{7/2}-2\times10^{-20} r^{11/2}-50 r^{13}-8. r^{12}-r^{11}-0.1 r^{10}-2\times10^{-17} r^9+4\times10^{-18} r^8$, $k_{13}=+2\times10^{-18} r^7+2\times10^{-19} r^6-3\times10^{-35} r^5-2\times10^{-36} r^4+7\times10^{-52} r^3-54 \sqrt{r}-4\times 10^{10}$.} & $1.5M$ & $1.77M$ & $\dfrac{\sqrt{r} \left(-0.3 r^{3/2}+r^3+4.5 r^2+6.75 r-0.9 \sqrt{r}+3.375\right)-0.68}{(r (r+3.)+2.25) (r (r (r+3)+2.25)-0.09)}$ & $\dfrac{k_{12}+k_{13}}{k_{11}}$\\
\hline
10  & $M$ & $M$ & $\dfrac{0.6}{r^3}$ &  $0$\\
\ChangeRT{1pt}
\end{tabular}}
\end{center}
\end{table*}

\begin{figure*}[th!]
\centering
\hbox{\includegraphics[scale=0.4]{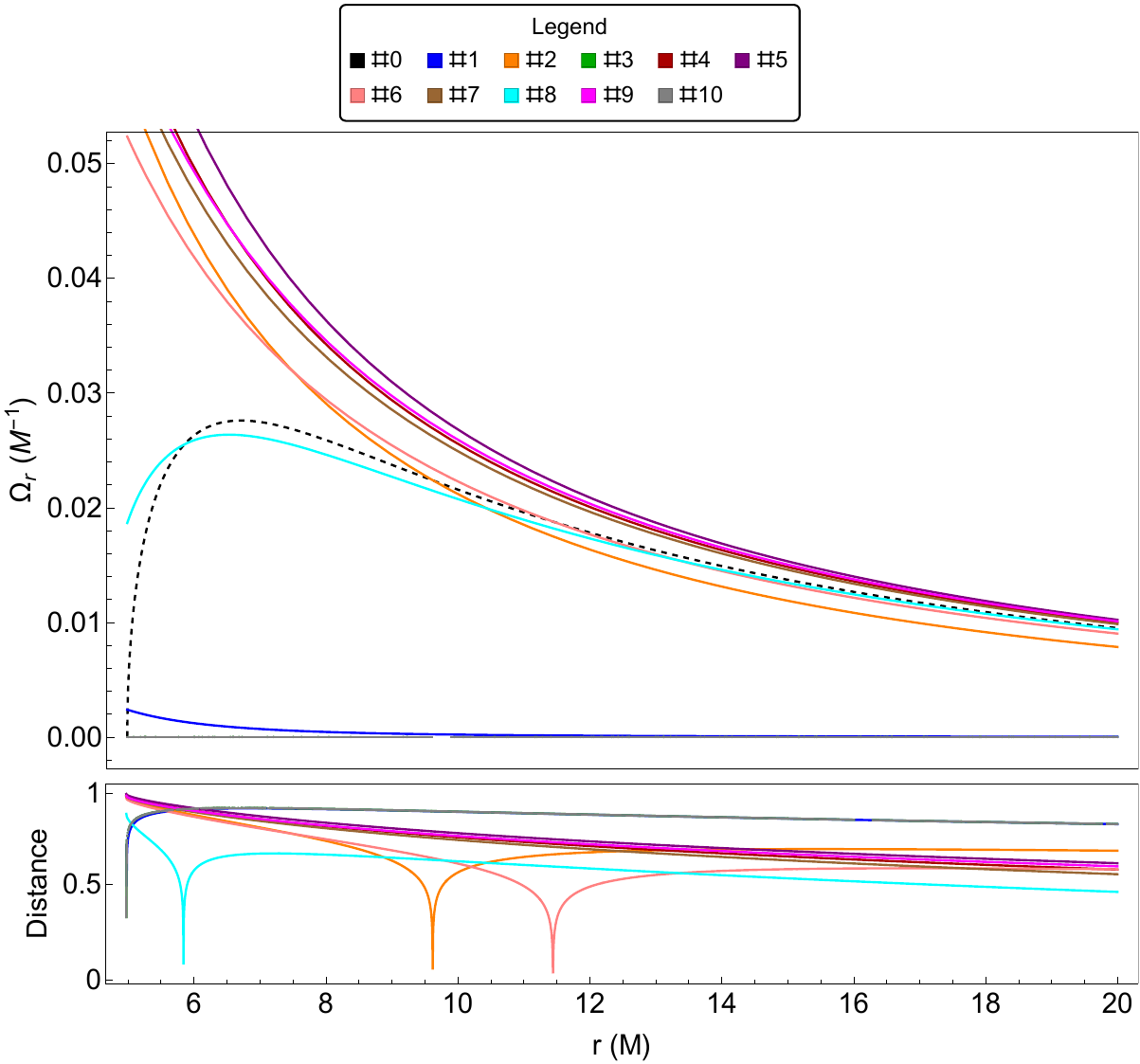}
\hspace{0.5cm}
\includegraphics[scale=0.4]{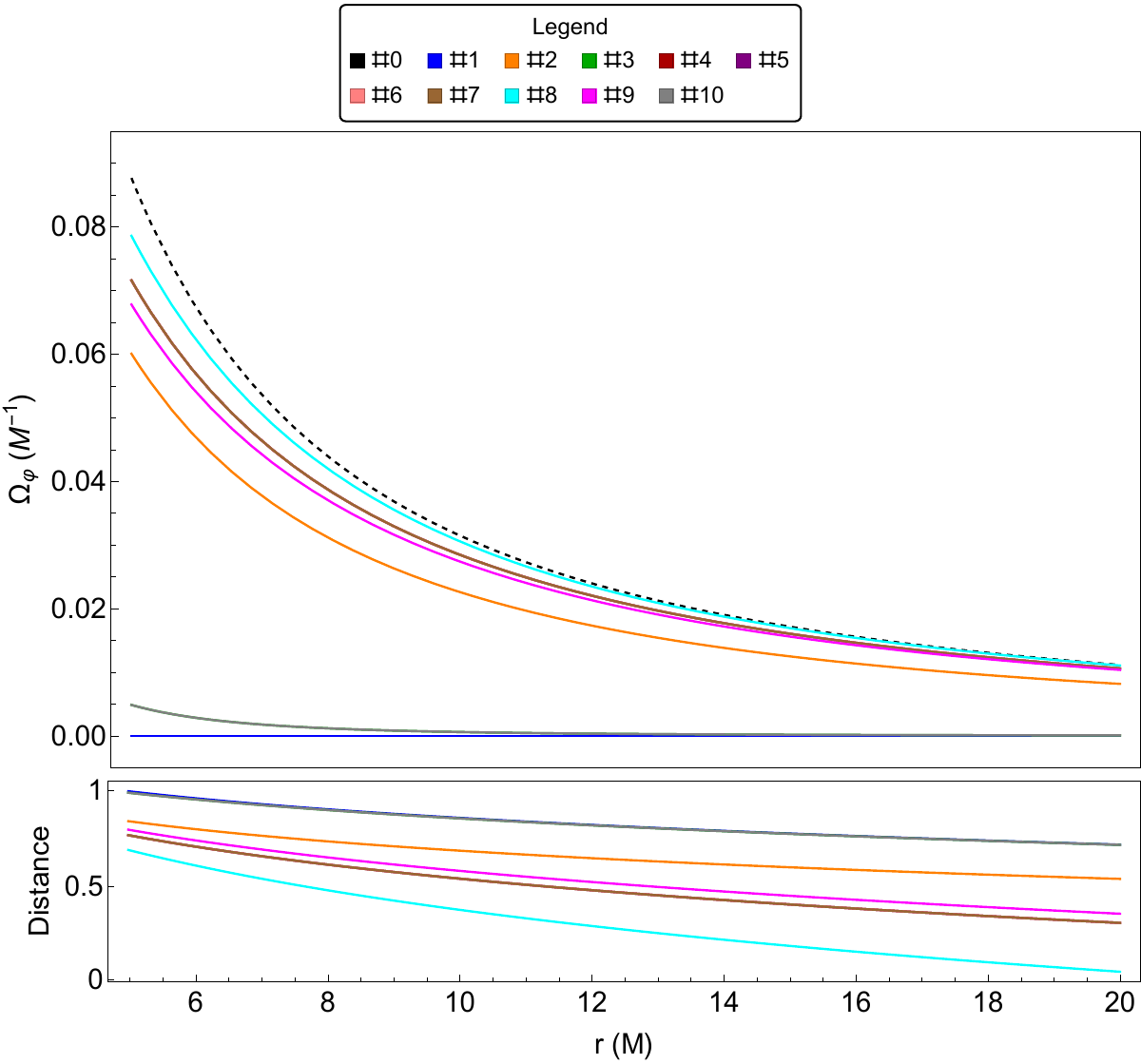}}
\caption{Plots of the angular epicyclic frequencies $\Omega_r$ (left panel) and $\Omega_\varphi$ (right panel) taken from Table \ref{tab:Table2}. The \qm{distance} in the lower panels of each figure represents the relative difference in absolute value between the angular epicyclic frequency of the WH solution (i.e., $\Omega_r^{\rm WH},\Omega_\varphi^{\rm WH}$) and the Kerr BH (i.e., $\Omega_r^{\rm BH},\Omega_\varphi^{\rm BH}$) expressed in normalized units. In other words, we have $|\Omega_r^{\rm WH} - \Omega_r^{\rm BH}|$ (left panel) and $|\Omega_\varphi^{\rm WH} - \Omega_\varphi^{\rm BH}|$ (right panel). The dashed black lines in all figures are related to the Kerr BH case.}
\label{fig:Fig1}
\end{figure*}

To distinguish between a Kerr BH and Teo-like WH, we plot in Fig. \ref{fig:Fig1} the epicyclic angular velocities of the WH solutions reported in Table \ref{tab:Table2}. The following comments are in order. For $\Omega_r$, we see that all the WH solutions exhibit the same trend for $r\gtrsim10M$, due to the asymptotically flat condition; whereas for $r\lesssim10M$, it is evident the presence of large deflection from the GR case, with particular relevance around the Kerr ISCO radius. We note that the WH solutions $\#1$ and $\#10$ behave not adequately for all $r$-range, while the WH geometry $\#8$ is an example of a BH mimicker solution. We can eventually claim that measurements around the ISCO radius are fundamental to identify possible metric-departures and thus hints for the possible existence of WHs.  

Instead, looking at the $\Omega_\varphi$ profiles, we immediately recognize that all WH solutions, except $\#1$ and $\#10$, behave similarly. Also in this case, the only way to catch a WH solution can be performed via analyses carried out around the Kerr ISCO radius. Although this astrophysical method is very efficient, there could be the unfortunate case, where slight deviations from BH solutions in GR may occur, but the described procedure may fail in its objective. In this situation, alternative astrophysical methods must be employed. However, the last eventuality should be always taken into account in order to robustly cross check the achieved results.    

\subsubsection{Digression on our methodology and simulations}
\label{sec:remark_sim}
This section is devoted to better illustrate the methodology we have pursued in the previous section in order to distinguish between a Kerr BH and a Teo-like WH, as well as to clarify some aspects of the simulations showed in Fig. \ref{fig:Fig1}, based on the values reported in Table \ref{tab:Table1}.

It is important to note that we have chosen the free parameters of each WH solution in a way they could mimic as much as possible the Kerr BH geometry. However, they have been calibrated and displayed only for the case of spin $a=0.3$. Therefore, it is spontaneous to question whether such a choice is valid also for other spin values. 

\begin{figure*}[th!]
\centering
\hbox{\includegraphics[scale=0.3]{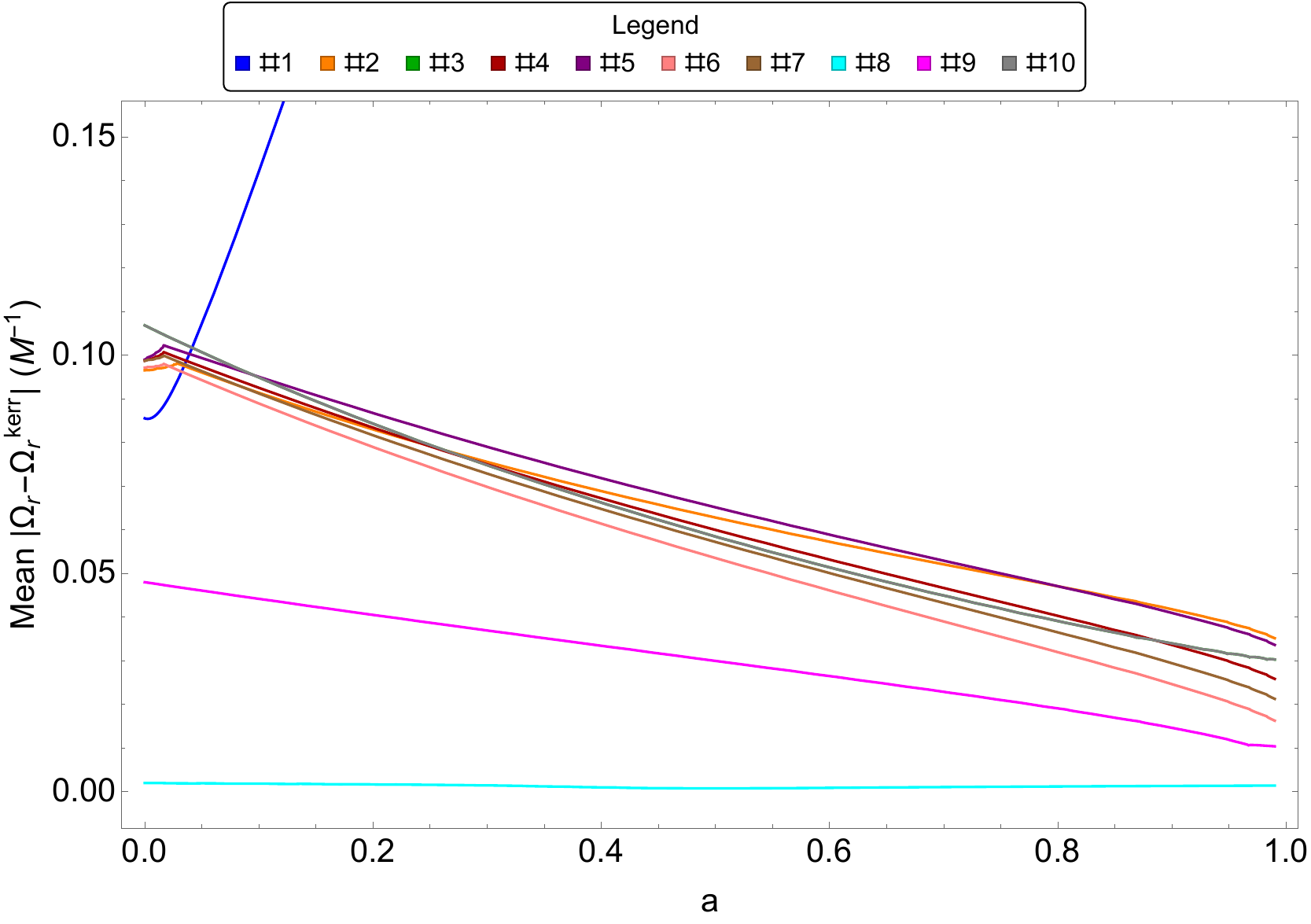}
\hspace{0.5cm}
\includegraphics[scale=0.3]{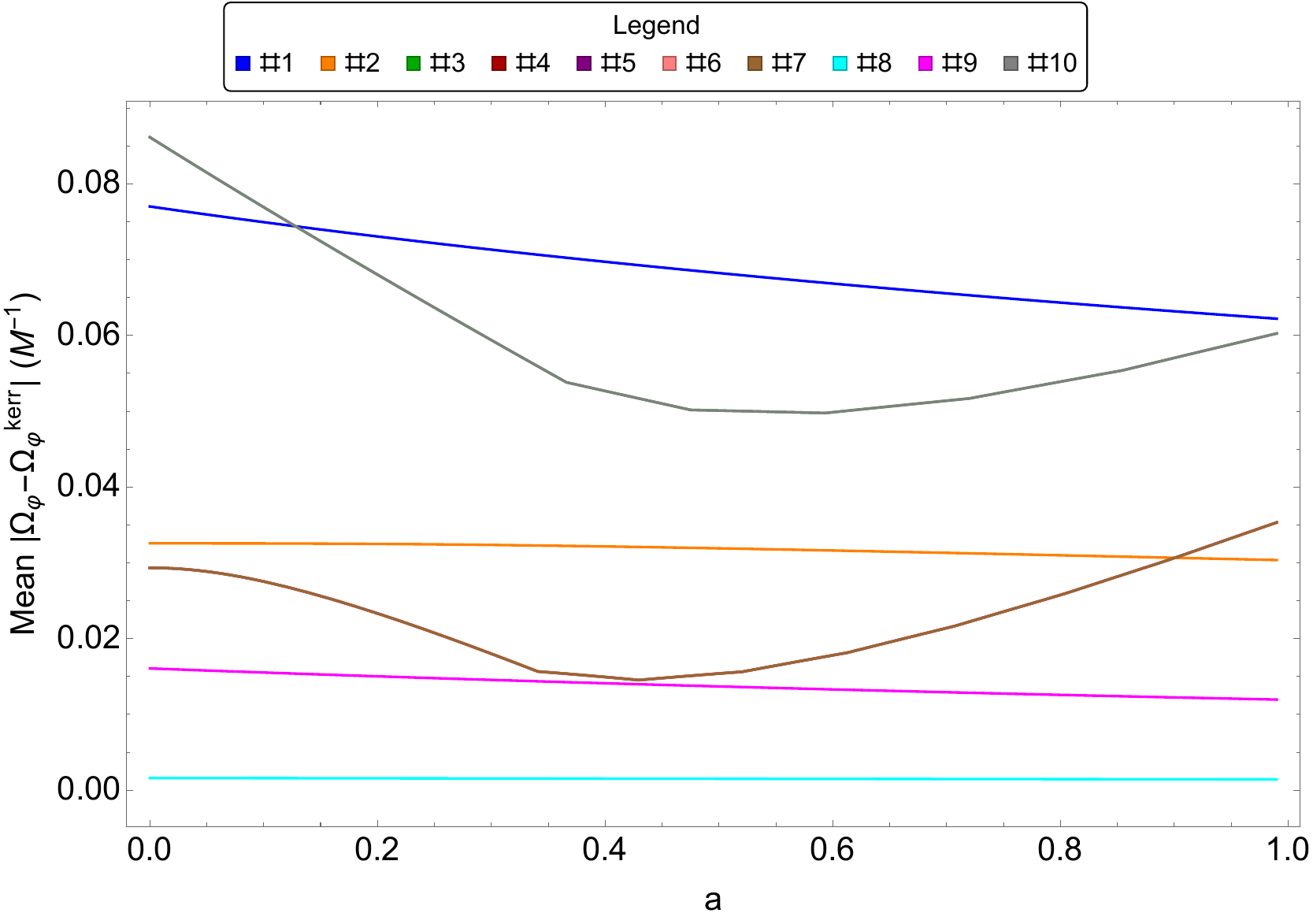}}
\caption{Plots of the mean absolute discrepancy between the WH angular epicyclic frequencies $\Omega_r$ (left panel) and $\Omega_\varphi$ (right panel) and the related functions $\Omega_r^{\rm Kerr}$ and $\Omega_\varphi^{\rm Kerr}$ for the Kerr BH, respectively (see Sec. \ref{sec:remark_sim}, for details). The considered WH solutions are always those reported in Table \ref{tab:Table1}, where all parameters have been fixed except for the spin value $a$.}
\label{fig:Fig2}
\end{figure*}

To this end, in Fig. \ref{fig:Fig2} we have produced two plots for both $\Omega_r$ and $\Omega_\varphi$. The procedure to realize them can be divided into three steps performed for each spin value $a$ and for each WH solution: (1) we calculate the WH ISCO radius, which is compared with that of the Kerr BH to finally select the minimum between them; (2) computation of the absolute discrepancy between the angular epicyclic frequencies of the WH solution and the Kerr BH, evaluated in 100 equally spaced points in the interval going from the appropriate ISCO radius (as explained in point (1)) to $20M$; (3) mean of the values collected in (2).

Figure \ref{fig:Fig2} is useful, because it constitutes a summary of the behaviour of the selected WH solutions in terms of the spin. However, if we generate plots similar to Fig. \ref{fig:Fig1} for some spin values covering the range $[0,1]$, we see that those WH solutions able to mimic the Kerr BH for $a=0.3$ fulfill the same job also for a generic value of $a$. We would like to underline that this situation is just a particular event. In the most general case, we would have changed the set of parameters for the selected WH solutions in order to mimic the Kerr BH solution for each fixed value of the spin. However, we underline that also in the worst case, our methodology does not fail in its objective. Indeed, from an astrophysical perspective when we focus on a gravitational system, this is determined by a precise value of the spin $a$. The present broad discussion, which contemplates disparate configurations, is finalized more on theoretically exploring how the selected WH solutions change in terms of their parameters.

\subsection{Reconstruction of wormhole solutions from the observational data}
\label{sec:WHrec}
Once a WH will be detected, it is fundamental to have a strategy, which permits to reconstruct the WH solution from the observational data. We have seen that in the static and spherically symmetric case, there is a balance between available equations (i.e., $\Omega_r$ and $\Omega_\varphi$) and unknown functions (i.e., $g_{tt}(r)$ and $g_{rr}(r)$) \cite{DeFalco2021EF}. Instead, in our case we have more unknown functions than available equations. In order to simplify the problem, we have already settled the metric in the equatorial plane $\theta=\pi/2$, entailing thus that the independent metric components are four, which depend only on $r$. Therefore, we need to complement the two constraints on epicyclic frequencies with two extra conditions.

Before to start, we adopt the following definitions:
\begin{subequations}
\begin{align}
g_{t\varphi}&=-r^2\omega(r),\\
g_{tt}&=-N^2(r)-g_{t\varphi}\omega(r),\\
g_{rr}&=\left[1-\frac{b(r)}{r}\right]^{-1},\\
g_{\varphi\varphi}&=r^2K^2(r).
\end{align}
\end{subequations}
We first present a procedure to reconstruct $\omega(r)$ (see Sec. \ref{sec:rec_omega}) and $K(r)$ (see Sec. \ref{sec:rec_K}) via some astrophysical techniques. Then, we are able to determine also $N(r)$ and $b(r)$ by exploiting the functions $\omega(r)$ and $K(r)$ and the data on the epicyclic frequencies (see Sec.  \ref{sec:rec_N_b}).

\subsubsection{Reconstruction of $\omega(r)$}
\label{sec:rec_omega}
Astrophysically, the data points on the $\omega(r)$ function can be acquired by measuring the frame-dragging effect at different radii. The sampled nodes could be gathered by adopting, for example, these strategies: line emission from an accretion disk \cite{Bromley1997,Ingram2012}, QPOs \cite{Cui1998}, and comparison between the numerical simulations of an accretion disk and the image provided by the Event Horizon Telescope (EHT) \cite{Ricarte2022}. Once, we collect them, we need to postulate a fitting function for reconstructing $\omega(r)$. To this end, it is useful to list some acceptable requirements: (1) $\omega(r)>0$ ($\omega(r)<0$) for positive (negative) values of $a$; (2) in modulus, it is a monotone decreasing function; (3) in the weak field limit, it behaves like $\omega(r)\approx 2Ma/r$ (as it also occurs in the Kerr metric). 

A reasonable and handy functional form of $\omega(r)$, meeting the aforementioned conditions, could be
\begin{equation} \label{eq:fit_omega_1}
\omega(r)=2Ma\left[\frac{r^\alpha+\sum\limits_{j=0}^{\alpha-1}a_j r^{j-1}}{r^{\alpha+1}+\sum\limits_{k=0}^\alpha b_k r^{k}}\right],
\end{equation}
where the coefficients $a_j$ and $b_k$ are real numbers (encoding the dependence from the WH mass, WH spin, and possibly other parameters).
We could assume that $\alpha\in[1,3]\subset \mathbb{N}$, taking inspiration from the WH solutions provided in Table \ref{tab:Table1}. Let us choose as general form that for  $\alpha=3$, which contains seven free parameters. A simpler way to reduce the complexity of the problem could be to Taylor-expand Eq. \eqref{eq:fit_omega_1} for $r\to\infty$, having thus
\begin{equation} \label{eq:fit_omega_2}
\omega(r)=2Ma\left[\frac{1}{r}+\frac{c_0}{r^2}+\frac{c_2}{r^3}\right],
\end{equation}
which involves only two free parameters (i.e., $c_0,c_1$). A strategy could be to first fit the data with Eq. \eqref{eq:fit_omega_2} in order to have a first rough estimation. Then, the analysis could be refined by employing Eq. \eqref{eq:fit_omega_1}.

We clarify that for the lack of observational data, we are just assuming a functional form of $\omega$. Equation \eqref{eq:fit_omega_2} contains six parameters, which can be further reduced depending on the available data and how they distribute. For example, one can truncate the series to lower orders or fixing some coefficients to certain numerical values. However, in this theoretical speculation we prefer to keep the general form, which could be further handled. 

\subsubsection{Reconstruction of $K(r)$}
\label{sec:rec_K}
The scheme to reconstruct $K(r)$ is more complicate, because this function does not have a direct physical effect as $\omega(r)$. In this case, it is not straightforward to determine the data points to be fitted. However, they  could be constructed as follows: (1) measuring some proper radial distances $\{R_i\}_{i=1}^\mathcal{M}$ with $\mathcal{M}>1$, like for example the photonsphere, the ISCO radius, and other regions obtained via the techniques already outlined for $\omega(r)$ \cite{Bromley1997,Ingram2012,Cui1998,Ricarte2022}; (2) we can associate to each distance $R_i$ from point (1) the related radius $r_i$, obtained by considering that the compact object is described by the Schwarzschild metric, whose mass can be estimated already at point (1); (3) we gather together the steps carried out in (1) and (2) to eventually build up the nodes $\{r_i,K_i\equiv R_i/r_i\}_{i=1}^\mathcal{M}$. 

For wisely restricting the functional space to search for $K(r)$, we remind that it is related to the proper radial distance $R(r,\theta)=rK(r,\theta)$ and must fulfill the following properties: (1) since $R(r)>0$ and $r>0$, we have  $K>0$; (2) $K(r)\to 1$ for $r\to+\infty$; (3) $K(r)$ must be finite, positive, and monotone decreasing everywhere outside the WH throat; (4) $\partial R(r)/\partial r>0$ implies $0>K'(r)>-K(r)/r$, where from now on the prime will stay for the derivative with respect to radial coordinate $r$. We emulate the functional form of the $K(r)$ function from the Kerr metric, which reads as
 \begin{equation}
K_{\rm Kerr}(r)=\sqrt{\frac{r^3+a^2(r+2M)}{r^3}}.
 \end{equation}
Indeed, we hypothesize that a possible general form of $K(r)$ for Teo-like WHs could be
 \begin{equation} \label{eq:KWH}
K(r)=\left(\frac{r^\beta+d_0}{r^\beta+d_1}+\sum_{i=0}^\mathcal{N}\frac{e_i}{r^{\beta+1+i}}\right)^{1/(2\gamma)}.
 \end{equation}
This expression has $\mathcal{N}+5$ free parameters, namely $\{\beta,d_0,d_1,\gamma,e_0,\dots,e_\mathcal{N}\}$ with $d_0>d_1$, $\beta>1$, and $\gamma\geq1$. Equation \eqref{eq:KWH} can be further simplified by setting: $\gamma=1$ to reduce the complexity of the ensuing fitting function and also the number of free parameters; $\mathcal{N}=2$, because higher-order terms strongly decrease, giving just tiny contributions. These further assumptions entail 
 \begin{equation} \label{eq:KWH2}
K(r)=\left(\frac{r^\beta+d_0}{r^\beta+d_1}+\frac{e_0}{r^{\beta+1}}+\frac{e_1}{r^{\beta+2}}+\frac{e_2}{r^{\beta+3}}\right)^{1/2}.
 \end{equation}
In this way, we are left with only six free parameters. A further helpful simplification could be in considering an asymptotic expansion of Eq. \eqref{eq:KWH2}, namely 
 \begin{equation} \label{eq:KWH3}
K(r)=1+\frac{A}{r}+\frac{B}{r^2}+\frac{C}{r^3},
 \end{equation}
 where we reduce to four free parameters. We use the same approach devised for $\omega(r)$, namely we first fit the data via the function \eqref{eq:KWH3}. Then, we ameliorate our analysis by exploiting Eq. \eqref{eq:KWH2} to obtain more precise results. 

As discussed at the end of Sec. \ref{sec:rec_omega}, also in this case, we prefer to keep the general form of Eq. \eqref{eq:KWH2}, which could be useful for eventual further manipulations.

\subsubsection{Reconstruction of $N(r)$ and $b(r)$}
\label{sec:rec_N_b}
Once we know $\omega(r)$ and $K(r)$, we are able to reconstruct $N(r)$ and $b(r)$ via the epicyclic angular velocities $\left\{\Omega_\varphi,\Omega_r\right\}$. We assume they are sampled in $n$ values $\left\{\bar{x}_i\right\}_{i=1}^n$ contained in the interval $[r_1,r_2]$, which we split in $n+1$ equally spaced points (i.e., $r_1\equiv x_0<\dots<x_{n-1}<x_{n}\equiv r_2$) such that $\bar{x}_i\in[x_{i-1},x_i]$ for every $i=1,\dots,n$. 

Therefore, from Eq. \eqref{eq:Omega_f} we obtain
\begin{align}
\left(\frac{N^2}{2}\right)'&=\frac{(g_{\varphi\varphi}')^2\Omega_\varphi^2+2g_{t\varphi}'g_{\varphi\varphi}'\Omega_\varphi}{2g_{\varphi\varphi}'}\notag\\
&-\left(\frac{g_{t\varphi}'\omega+g_{t\varphi}\omega'}{2}\right).
\end{align}
Discretizing this equation, we have
\begin{align}
\frac{N^2(\bar{x}_i)}{2}&=\frac{N^2(r_1)}{2}+\sum_{i=1}^N\biggr{[}\frac{(g_{\varphi\varphi}'(\bar{x}_i))^2\Omega_\varphi^2(\bar{x}_i)}{2g_{\varphi\varphi}'(\bar{x}_i)}\notag\\
&+\frac{2g_{t\varphi}'(\bar{x}_i)g_{\varphi\varphi}'(\bar{x}_i)\Omega_\varphi(\bar{x}_i)}{2g_{\varphi\varphi}'(\bar{x}_i)}-\frac{g_{t\varphi}'(\bar{x}_i)\omega(\bar{x}_i)}{2}\notag\\
&-\frac{g_{t\varphi}(\bar{x}_i)\omega'(\bar{x}_i)}{2}\biggr{]}(x_i-x_{i-1}).
\end{align}
The above expression can be entirely calculated, since we know the functional form of $\omega(r)$ and $K(r)$. The only unknown value is $N(r_1)$, which can be estimated by following the same scheme devised in Ref. \cite{DeFalco2021EF}. From this first step, we have the following points $\{\bar{x}_i,N(\bar{x}_i)\}_{i=1}^n$ to be fitted in order to reconstruct $N(r)$.

From Eq. \eqref{eq:Omega_r}, we have 
\begin{subequations} 
\begin{align} 
Z&=\frac{(g_{tt}+\Omega_\varphi g_{t\varphi})^2}{2}\biggr{[}\partial_{rr}^2\biggr{(}\frac{g_{\varphi\varphi}}{Y}\biggr{)}+2\ell \partial_{rr}^2\biggr{(}\frac{g_{t\varphi}}{Y}\biggr{)}\notag\\
& \hspace{3cm}+\ell^2\partial_{rr}^2\biggr{(}\frac{g_{tt}}{Y}\biggr{)}\biggr{]},\\
b(\bar{x}_i)&=\sum_{i=1}^N \bar{x}_i\left(1-\frac{\Omega_r(\bar{x}_i)}{Z(\bar{x}_i)}\right).
\end{align} 
\end{subequations} 
By fitting the points $\{\bar{x}_i,b(\bar{x}_i)\}_{i=1}^N$, we reconstruct $b(r)$.

In this case, we do not provide some general expressions for both the fitting functions, since they can be, in general, of any form. In addition, they are important for characterizing the WH solution and the gravity theory from which they come. Therefore, we list only some general constraints, which these functions must fulfill: 
\begin{enumerate}
\item $N(r)$ must be a positive monotone decreasing function, which asymptotically tends to 1. For recovering the Newtonian theory in the weak field limit, we have that for large radii $N(r)\to\sqrt{1-2M/r}$;
\item $b(r)$ must be a positive monotone increasing function such that $b(r)< r$ and asymptotically it should behave like $b(r)/r\to0$. Finally, the flaring out condition imposes that the derivative must satisfy $b'(r)<b(r)/r<1$. For recovering the Newtonian theory in the weak field limit, we have that for large radii $b(r)\to2M$.
\end{enumerate}

\subsection{Conclusions}
\label{sec:end}
In this paper, we have considered the epicyclic frequencies in the equatorial plane around general stationary, axially symmetric, and traversable WHs, modeled by the Teo-like metric. We have first described the general properties of this class of WHs and then we have written the formulas of the epicyclic frequencies in terms of the metric components (see Sec. \ref{sec:epicyclic-frequencies-Teo-WHs}). Subsequently, we have used the formulas of the epicyclic frequencies for detecting the eventual presence of a WH (see Sec. \ref{sec:WHdet}). Since we do not have yet data on WHs, we have considered some WH solutions proposed in the literature, see Table \ref{tab:Table1}. For each WH, we have also calculated the WH throat $r_0$ and the ISCO radius $r_{\rm ISCO}$, as well as the explicit expressions (in the equatorial plane) of $\Omega_\varphi$ and $\Omega_r$ (once the free parameters have been fixed), see Table \ref{tab:Table2}. In Fig. \ref{fig:Fig1} we have shown the profiles of the epicyclic frequencies compared to those obtained in the Kerr metric. From these plots we deduce that analyses carried out around the Kerr ISCO radius are fundamental to highlight possible metric-departures from GR. Our study has been carried out for a fixed value of the spin. However, we have verified also that by changing the spin values, the selected WH solutions behave similarly to the displayed case (see Fig. \ref{fig:Fig2} and Sec. \ref{sec:remark_sim}, for details).

Finally, in Sec. \ref{sec:WHrec} we present a strategy to reconstruct the WH solution once the observational data on WHs will be available. Since there are four unknowns $\{N(r),b(r),\omega(r),K(r)\}$ and only two equations $\{\Omega_\varphi,\Omega_r\}$, we need two extra constraints. We propose some procedure to reconstruct $\omega(r)$ and $K(r)$. Regarding the function $\omega(r)$, we first construct the observational data via the measurement of the frame-dragging effect in some radii and then we fit them via some selected functions (see Sec. \ref{sec:rec_omega}). Instead, for the function $K(r)$ the reconstruction process is more complex, especially for the assembly of the observational data. Also in this case, we are able to select some general functional forms of $K(r)$ for fitting the data (see Sec. \ref{sec:rec_K}). In the last part, we use the data on epicyclic frequencies and the explicit expressions of $\Omega_\varphi(r)$ and $\Omega_r(r)$, together with the analytical expressions $\omega(r)$ and $K(r)$, to reconstruct also $N(r)$ and $b(r)$. We would like to stress that we have proposed a general strategy, which could be improved in terms not only of the construction of the data, but also in terms of the fitting functions (depending on the given nodes).

This work can be applied not only to WHs, but also to investigate other compact objects. Furthermore, the capacity to detect metric-departures around the ISCO radius is extremely important for providing tests of gravity within GR or Extended Theories of gravity. In particular, the epicyclic frequencies permits to easily reconstruct from the data either a WH metric or also a BH solution framed in another gravity theory different from GR. As remarked also in this paper, sometimes it could be difficult to detect the WH solution or to reconstruct its metric by only exploiting the epicyclic frequencies. Therefore, it is always useful to complement this approach with other astrophysical methods in order to have more solid results.   

As future perspectives, we aim at extending this strategy to the whole three-dimensional space around stationary and axially symmetric WH geometries, where the role of the polar epicyclic angular velocity becomes extremely useful. We envisage the following issues to be addressed: (1) the Teo metric \eqref{eq:Teo_WH} must be modified, because there should be five unknown functions (in correspondence with the five metric components) in order to faithfully model WHs in the three-dimensional space; (2) determining $\Omega_\theta(r,\theta)$ outside the equatorial plane (whose formula is not coincident with Eq. \eqref{eq:Omega_th}, valid only in the equatorial plane); (3) all metric components will be functions of $(r,\theta)$, meaning that we need to find samples along the $r$ and $\theta$ directions; (4) the fitting procedures will occur in the three-dimensional space, where the nodes must be interpolated by two-dimensional surfaces.    

\section*{Acknowledgements}
V.D.F. thanks Gruppo Nazionale di Fisica Matematica of Istituto Nazionale di Alta Matematica for the support. V.D.F. acknowledges the support of INFN {\it sez. di Napoli}, {\it iniziative specifiche} TEONGRAV.

\bibliography{references}

\end{document}